\documentclass[12pt,preprint]{aastex}
\newcommand\kms{km~s$^{-1}$\ }

\shorttitle{BAL Quasars in the SDSS}
\shortauthors{Tolea, Krolik, and Tsvetanov}

\begin{document}

\title{Broad Absorption Line Quasars in the Early Data Release from the
Sloan Digital Sky Survey}

\author{Alin Tolea, Julian H. Krolik, Zlatan Tsvetanov,}

\affil{Department of Physics and Astronomy, Johns Hopkins University,
    Baltimore, MD 21218}



\begin{abstract}

A new broad absorption line quasar (BAL) sample is derived from the
first data released by the Sloan Digital Sky Survey.  With 116
objects, it is the largest BAL sample yet assembled.  Over the
redshift range $1.8 \leq z \leq 3.8$, the crude fraction with broad
absorption in the \ion{C}{4} line is $\simeq 15\%$.  This fraction
may be subject to small selection-efficiency adjustments.  There
are also hints of redshift-dependence in the BAL fraction.
The sample is large enough to permit the first
estimate of the distribution of ``balnicity index": subject to certain
arbitrary parameters in the definition of this quantity, it is very
broad, with (roughly) equal numbers of objects per logarithmic
interval of balnicity.  BAL quasars are also
found to be redder on average than non-BAL quasars.  The
fraction of radio-loud BAL quasars is (weakly) consistent with
the fraction of radio-loud ordinary quasars.

\end{abstract}


\section{Introduction}

     Broad Absorption Line quasars (BALs) are one of the most
enigmatic varieties of quasars.  Resonance lines of ordinary
ions---\ion{H}{1}, \ion{C}{4}, \ion{N}{5}, \ion{O}{6}, \ion{Mg}{2},
and others---are seen in absorption that spreads, often in highly
irregular fashion, as much as 60,000~\kms from line-center in the
quasar rest-frame to the blueward.  Previous surveys (e.g. the
Large Bright Quasar Survey, or LBQS:
\citet{W91}) have shown that BALs, while a minority of all quasars,
are not rare; a population fraction $\sim 10\%$ is typically
estimated.  Because few of their other properties are grossly
different from ordinary quasars, it is generally thought that all
quasars have BAL material, but it covers only a fraction of solid
angle around the quasar nucleus \citep{W91}.  However, subtleties
of selection can complicate
the inference of covering fraction from population fraction
\citep{G97,KV98}.

     Numerous technical difficulties have retarded growth in our
understanding of BALs.  Known cases are relatively rare, numbering
less than $\sim 100$, not solely because they are a minority of the
general population but also because they are readily found only when
their characteristic features are red-shifted from their rest-frame
wavelengths in the ultraviolet into the visible band.  Consequently,
only those quasars found in somewhat special redshift intervals can be
easily searched for broad absorption.  It is hard to statistically
characterize those BALs that are found because the methods used to
discover them often involve some level of subjectivity that is hard to
quantify.  Even if their selection were easier to articulate, there
appears to be so much variation in their properties (profile shapes,
relative line strengths, etc.) that it is hard to grasp which
properties are generic and which are ``accidental".

    The quasar sample being compiled by the Sloan Digital Sky Survey (SDSS:
\citet{Y00}) offers a way out of this impasse.  When complete, it will be
both very large ($\sim 10^5$ in all) and selected in a uniform and
quantifiable manner.  In future work, we hope to present statistical
analysis of BALs in this entire sample.  Here we offer a
preliminary installment on this project in the form of a more modest
BAL sample drawn from the first data released from the project to
public view, the Early Data Release (EDR: \citet{Sto02}).

    Several collections of BALs have already been drawn from early Sloan
data \citep{M01,H02}; these were, however, oriented toward ``by-eye"
selection of small subsamples special in some way (radio-loud in the
former case, extraordinary profiles in the latter).   The work reported
here differs in that it is the first attempt to create a
systematically-selected sample from the SDSS.

      From the EDR, \citet{Sch02} created a quasar catalog containing
3814 quasars, selected (mostly) on the basis of their location in
four-color space and on a (mostly) uniform $i$-magnitude limit.  In
order to present more clearly-defined statistics, we have refined
this sample so that it is almost homogeneously-selected (see \S 2.1).
Within that sub-sample (about 80\% of the full EDR quasar catalog), roughly
one-quarter (796) fall within the redshift range within which it is feasible
to search for \ion{C}{4} BAL features.

      With an eye toward the homogeneity of selection to be achieved in
the full SDSS, we invented an automated BAL selection algorithm that
processes SDSS spectral data in a
uniform way and identifies BAL quasars in a uniform manner (see
\S 2.2).  Using this algorithm, we have identified 116 BAL quasars,
whose statistical properties are discussed in \S 3.  Although the
EDR represents a tiny fraction of the ultimate SDSS quasar sample, the
BAL sample so derived is now the largest (as well as the most
homogeneously selected) such sample known. 

\section{Details of Sample Selection}

\subsection{Quasar selection}

     The EDR quasar catalog \citep{Sch02} was compiled by applying
several different selection criteria (see \citet{Sto02} for details).
Most of its objects were chosen on the basis of colors lying
outside the ``stellar locus" in the four-color space formed
by the Sloan five-filter ($u$, $g$, $r$, $i$, and $z$)
photometry\footnote{Technically, because the photometry published in the
EDR had not received final calibration, the magnitudes were shown as
$i^*$, etc.  In this paper, all such magnitudes should be understood
in that sense, although we will forgo making the distinction
explicit.}.  However, roughly 20\% of the quasars in
this list were chosen in a much less well-defined fashion (in the
jargon of the SDSS, these were selected for the ``serendipity" sample).
In addition, much smaller
numbers of quasars were selected not on the basis of their photometric
colors, but because they were close in the sky to known radio sources
in the FIRST catalog \citep{B95} or X-ray sources in the Rosat All-Sky Survey
\citep{V99}.  For the purposes of this paper, which concentrates on
statistics, we have pruned the EDR quasar catalog to include {\it only} those
flagged as quasar candidates by a color-based targetting algorithm.

    Because the EDR data were compiled during the test-year of the project,
the color-based selection algorithm was not the same throughout.  With
regard to issues relevant here, the variations can be reduced to
two slightly different versions, very nearly equal in sky coverage.  In
both, the primary flux limit was $i \le 19$~mag
\footnote{Objects were also required to be fainter than
a limit chosen to avoid image saturation and cross-talk between
spectroscopic fibers.  This was $i = 15.0$~mag in some
cases, $i=16.5$~mag in others; because so few objects are near the
bright limit, the change makes essentially no difference to
sample statistics.}.
Both versions also shared the same primary color criterion: select only
those objects whose colors lie at least $4\sigma$ away from the
stellar locus.  An exception was made to this rule in order
to cope with the fact that at redshifts $z \simeq 2.5$ -- 3.0, quasars
have colors nearly indistinguishable from those of A and/or F
type stars \citep{F99,R01}.
In this portion of the stellar locus, quasar candidates were selected, but
at lower efficiency (Stoughton et al. 2002; G.T. Richards, private
communication).

   The two versions differed in two ways:  One version rejected
objects with colors approximating those of A stars;
in the other, objects with colors similar to those of hot white dwarfs or
unresolved M dwarf-white dwarf pairs were also removed from the
quasar candidate list.  In addition, in those runs in which white-dwarf-like
colors were rejected, a special procedure was adopted in order to enhance
sensitivity to quasars with $z \geq 3.5$.   To find more of these quasars, the
magnitude limit was relaxed to $i=20$ in the
region of four-color space where previously-located high-redshift quasars
were found.
We will show below that these variations in quasar candidate selection had
negligible effects on BAL quasar discovery.

    Quasar candidates were labelled as ``quasars" by the spectroscopic
pipeline if the crosscorrelation between their spectra and a quasar
template spectrum was greater than the crosscorrelation with any of
the other templates (stars, galaxies, etc.).  For confirmation, objects
were required to pass two further tests: that their spectra possess at
least one emission line with FWHM $ \geq 1000$~km~s$^{-1}$; and that
their absolute magnitude $M_i \leq -23$ (for $H_0 = 50$~km~s$^{-1}$ and
$q_0 = 0.5$).

   In the full EDR quasar catalog, there were 3814 objects.  Our sample
contains only the 3107 identified by one of the two versions of the
color-selection rules.

\subsection{BAL identification}

    The \ion{C}{4} line is centered at 1550~\AA\ in the rest-frame, so it
appears in the SDSS
spectra (which nominally cover the wavelength range from 3900--9100~\AA)
only for redshifts $1.5 < z < 4.9$.  However, several effects limit this
range further.  First, although the nominal blue cut-off is 3900~\AA,
in practice, throughput, and therefore signal/noise, drop sharply shortward
of $\simeq 4100$~\AA\ and (more gradually) longward of $\simeq 8000$~\AA.
Moreover, in order
to measure possible absorption to the blue of line-center, we must be
able to see some line-free continuum to the redward of the emission line-center
and also follow the line far enough to the blue that we are confident we
have defined the entire absorption profile.  These requirements 
restrict the permissible range of redshifts to roughly $1.8 \leq z \leq 3.8$,
cutting our sample size to 796 objects.

    Redshifts supplied by the SDSS spectroscopic pipeline are typically
accurate to $\sim 1000$~\kms, which
suffices for the cut described in the previous paragraph, but is not accurate
enough for absorption line measurement.  These measurements require
greater accuracy because the classical definition of BALs
\citep{W91} counts only absorption at least 3000~\kms to the
blue of line-center in the rest-frame.

    The \ion{C}{4} emission line cannot be used to define the redshift
to this level of accuracy because the very absorption we are interested
in can cut into the emission line so severely that it is unclear where
line-center occurs.  Instead, we define the quasar rest-frame in terms
of the \ion{C}{3}]~1909 emission line\footnote{According to \citet{VdB01},
this line is, on average, offset only $\simeq 200$~km~s$^{-1}$ from the
systemic host redshift as defined by \ion{O}{3}[5007].  For
similar reasons, \citet{W91}
used a weighted mean of the \ion{C}{4}, \ion{C}{3}, and \ion{Mg}{2}
emission lines to determine their redshifts.}.  To determine
the redshift this way, we fit a Gaussian plus a linear component to the
measured flux in the (pipeline-redshift) rest-frame wavelength range
1860 -- 1960~\AA.  The center of the Gaussian in the best fit we take to
define the true observed wavelength for rest-frame 1909~\AA.

     To search for absorption---which can be extremely difficult in these
objects, in which emission and absorption features can occupy most of the
spectrum---one must first locate the continuum.
Our solution to this problem is to first fit a power-law to the continuum
data in five line-free windows: 1790--1820~\AA, 1975--2000~\AA, 2140--2155~\AA,
2240--2255~\AA, and 2265--2695~\AA.  Holding that component fixed, we then
fit a half-Gaussian to the red half of the \ion{C}{4} emission feature
lying above the fitted continuum, taking line-center as 1549.5~\AA\ in
the rest-frame (this wavelength derived by equally weighting the
two components of this doublet).  We deliberately ignore the blue
half of the emission line so as to avoid confusion by absorption; the
continuum windows are chosen so as to avoid contamination by the
\ion{He}{2}~1640, \ion{C}{3}]~1909, and \ion{Mg}{2}~2800 emission lines,
as well as from various \ion{Fe}{2} emission complexes.  We then extrapolate
the power-law portion of this fit to define the continuum blueward of 1550~\AA.

    

      The final step in our procedure is to compute the ``balnicity index"
for each quasar in this redshift range, following the definition given in
\citet{W91}:
\begin{equation}
BI = \int_{-25,000}^{-3000} \, dv \, \left(1 -
          {F_\lambda \over 0.9 C_\lambda}\right) {\cal C},
\end{equation}
where the measured flux per unit wavelength is $F_\lambda$, the
extrapolated fitted continuum is $C_\lambda$, and ${\cal C}$
is a function whose value is unity when the quantity between
parentheses has been positive for at least 2000~\kms\ to the red
of the current wavelength and zero otherwise.  The lower-limit
on the integral is designed to avoid confusion with the \ion{Si}{4}
line, the upper-limit to exclude associated absorption.  Comparison
is made to 0.9 times the continuum rather than the full continuum
to be conservative with respect to noise features.  The function
${\cal C}$ ensures that only truly broad features are counted.  In
effect, the balnicity index amounts to a sort of equivalent width.
Following Weymann et al., we declare a quasar to be a BAL when its
balnicity index is greater than zero.

\section{Results}

   In view of the preliminary nature of this sample, here we present
only a few tentative results.  The statistical and systematic
uncertainties in the
numbers presented here will be substantially reduced in the far larger
full SDSS sample; that sample will also enable other, more detailed
studies.

\subsection{BAL fraction}

\begin{figure*}
\plotone{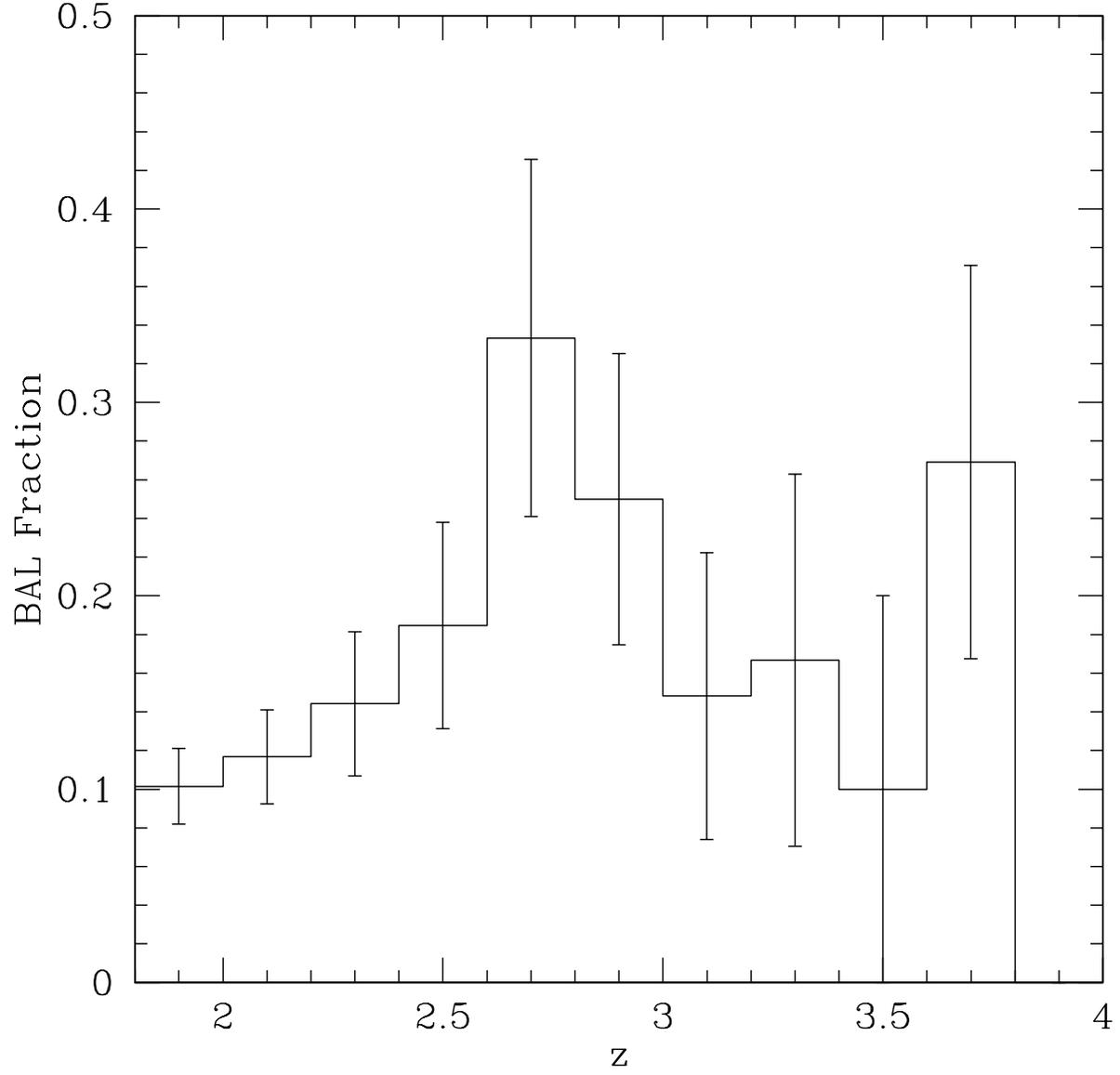}
\caption{BAL fraction in the sample as a function of redshift.  Errorbars
are $1 \sigma$, and purely statistical.
\label{balfrac}}
\end{figure*}

   Of the 796 quasars in the redshift range where we can search for
BALs, we find that 116, or $\simeq 15\%$ are \ion{C}{4} BALs.
The apparent BAL fraction in this sample is also strongly dependent upon
redshift (fig.~\ref{balfrac}): it is only $\simeq 10\%$ in the
redshift range $1.8 \leq z \leq 2.2$, but rises to $\simeq 33\%$
near $z \simeq 2.7$, and averages $\simeq 20\%$ for $2.2 \leq z \leq 3.8$.

     Some of this redshift dependence may be real, but there are also
redshift-dependent systematic effects.   Near $z \simeq 1.8$,
some BALs may be lost due to the relatively poor S/N
at the blue end of the spectrograph.  In the range $2.5 \leq z \leq 3$,
distinguishing quasars and stars by color becomes difficult.  Both
the broad absorption itself and intrinsic differences in continuum
shape (\S 3.3) can give BALs colors different from ordinary quasars;
their selection efficiencies can therefore differ significantly
in this redshift range.  The spike in the ``raw" BAL fraction near
$z \simeq 2.7$ may be the result of this differential selection
efficiency (see \citet{Rei03} for further exploration of this issue).

     On the other hand, dividing the sample according to the
color-selection procedure used, we find negligible differences.
Both BAL and ordinary quasar selection efficiencies were equal
to well within Poisson errors.

     The LBQS found a significantly smaller ``raw" fraction (9\%),
but Weymann et al. corrected this figure to $\simeq 12\%$  because
the BAL itself removed enough flux
that many BAL quasars dropped below the survey flux-limit.  In the SDSS,
by contrast, the flux-limit is applied in the $i$-band, near 8500~\AA .
Only for redshifts $\simeq 4$ would a \ion{C}{4} 1550 BAL influence the
$i$-band flux, and we do not even consider such high-redshift quasars in
the sample at hand.  Consequently, the SDSS BAL fraction needs to be corrected
for this effect only at very high redshift or for the special case of
``LoBALs", BAL quasars with absorption in the \ion{Mg}{2} 2800 line.   When
\ion{Mg}{2} absorption is present, the BAL would remove flux from the
$i$-band when $z \simeq 2.5$.

    Overall, then, particularly for $z \geq 2.2$, the SDSS appears to
find a somewhat larger BAL fraction than the LBQS.  Allowing for
the various systematic errors, our best preliminary estimate is
a BAL fraction $\simeq 15$ -- 20\% for this redshift range, but possibly
nearer 10 -- 12\% for $1.8 \leq z \leq 2.2$.  We expect these numbers to
be refined as the statistics improve and the systematic effects become better
understood.   Using the full SDSS quasar sample it should become
possible to search for genuine redshift- (or luminosity-) dependence
in the BAL fraction.


\subsection{Balnicity index distribution}

\begin{figure*}
\plotone{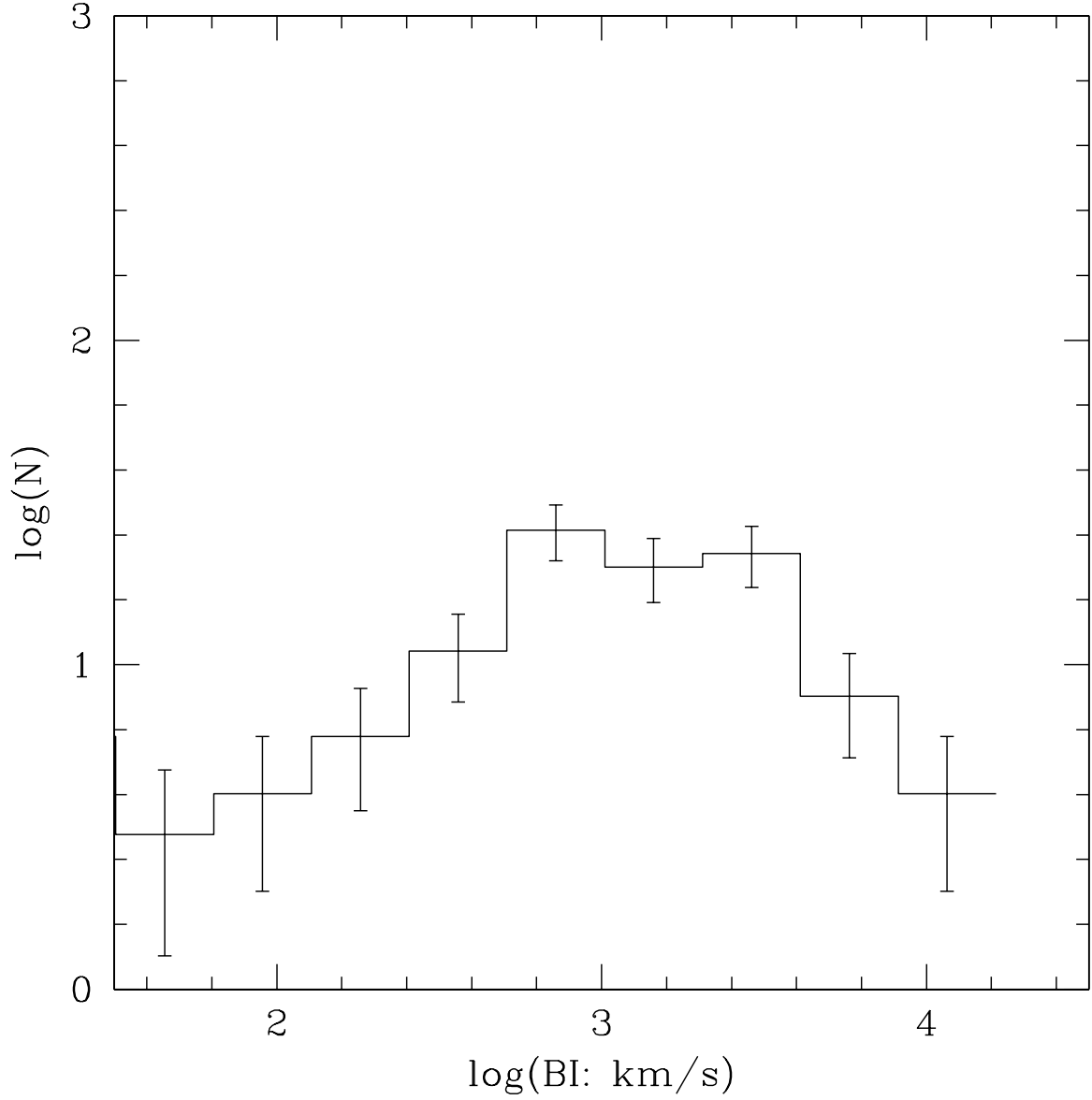}
\caption{
Distribution of BI in the sample (solid histogram).  The binning is
logarithmic in BI.  Errorbars are $1 \sigma$.
\label{bidist}}
\end{figure*}

    The BI distribution for our sample is shown in Figure~\ref{bidist}.
The plot shows $\log[dN/d\log(BI)] = \log[BI dN/d(BI)]$; within the
loose constraints placed by relatively small sample size, there are
roughly equal numbers of objects in equal logarithmic
bins.  However, we stress that the shape of this distribution below
$\sim 1000$~km~s$^{-1}$ is strongly dependent upon the arbitrary
velocity offset parameter used in the definition of balnicity to
distinguish ``associated absorbers" from truly ``broad" absorption
(cf. the discussion in \citet{H02}).

The shape of this distribution has several interesting implications.
First, because BALs of very small BI are common, the arbitrariness
of the velocity offset parameter means that the distinction between
weak BALs and ``associated absorbers" is difficult to mark and the
nominal BI for these objects likely underestimates the ``physical"
absorption.
Second, if we take the definition of balnicity at face value,
the breadth of its distribution is consistent with the
anecdotal sense of the diversity of BAL profiles derived from previous,
smaller samples.  
Third, the shallow slope of the distribution at the high-balnicity
end suggests that the maximum velocity width of BALs is probably as yet
ill-defined.

\subsection{Colors of BALs}
   
\begin{figure*}
\plottwo{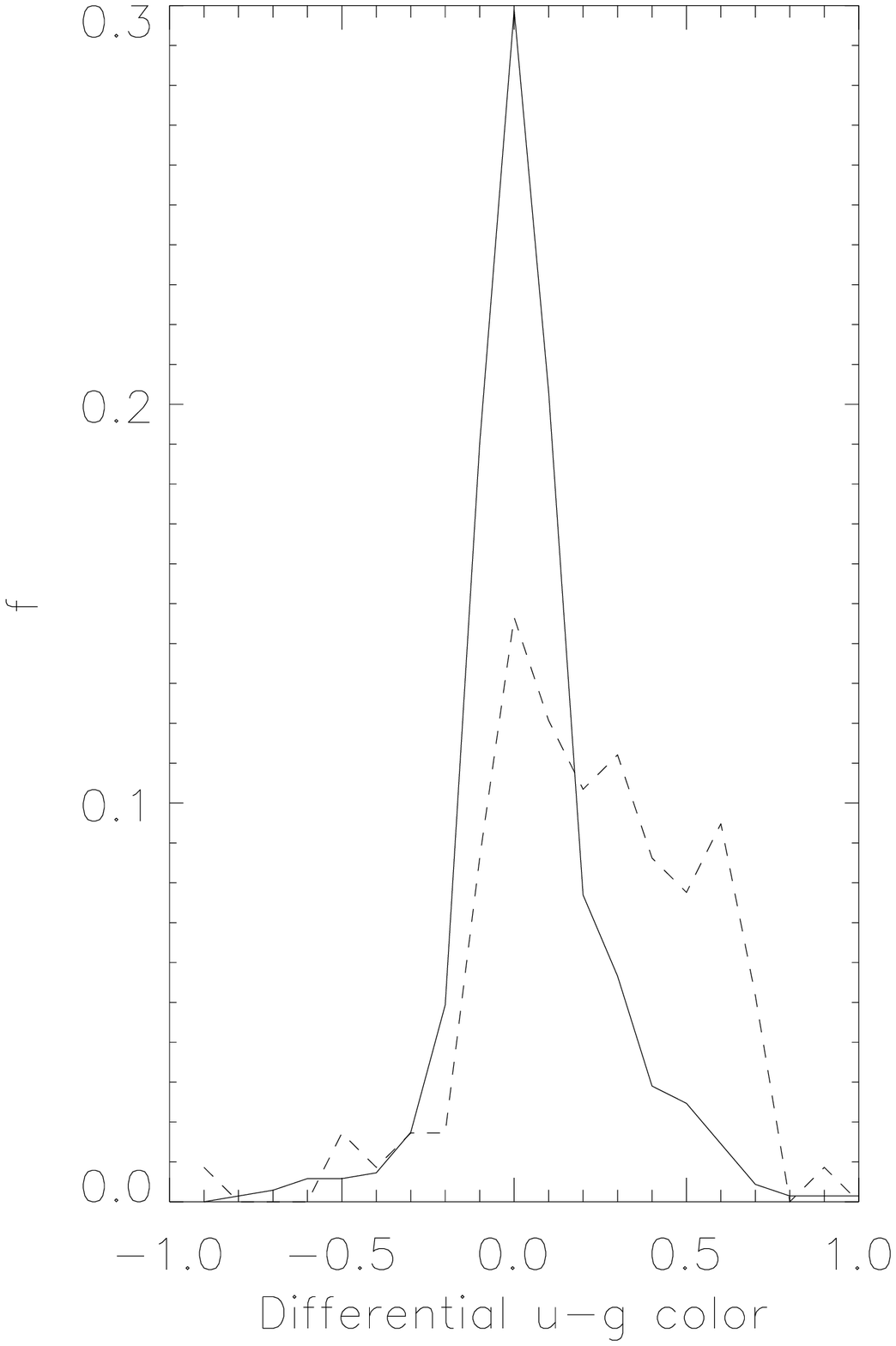}{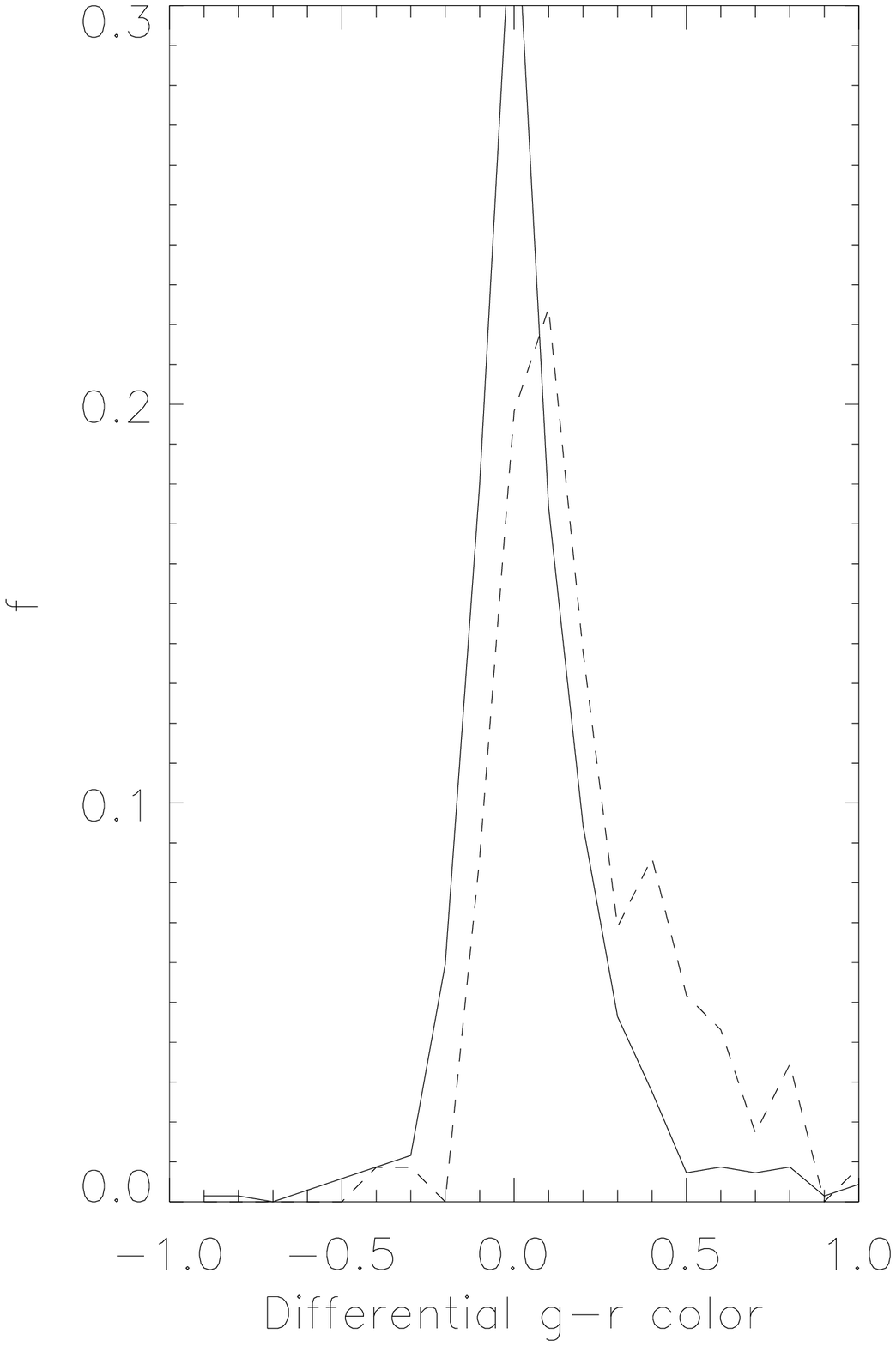}
\caption{The distribution of colors after subtracting off the mean color
of quasars at the individual quasars' redshifts.  Solid line is non-BAL
quasars, dashed line is BAL quasars.  The left-hand panel is the distribution
for normalized $u-g$; the right-hand panel for normalized $g-r$.
\label{BALcolors}}
\end{figure*}

   The mean quasar color is a strong function of redshift \citep{F99,R01}.
To contrast the colors of BALs and non-BAL quasars most clearly, in
Figure~\ref{BALcolors} we show the distribution of colors after subtracting
the mean color for our sample at each quasar's redshift.  Particularly
in $u-g$, the distribution of BAL colors is shifted distinctly to the red
(see also \citet{M01}).  In the mean,
the $g - r$ color difference is about 0.18~mag; in $u - g$ it
is about 0.34~mag.  Both color offsets are crudely constant with redshift
for $1.8 \leq z < 3$.  This trend is in the same sense
(although somewhat smaller than) the color contrast in the FIRST survey
\citep{B00,Br01}, in which radio-selected BALs were, on average,
$\simeq 0.5$~mag redder (in a color roughly equivalent to $B - R$) than
their radio-selected non-BAL quasars.

    There are several possible explanations for this effect.  It may be,
for example, that there is dust associated with the absorbing
material itself.  It is also possible that the redder colors are due
to dust, but farther from the nucleus along our line-of-sight (e.g.,
as proposed by \citet{G97}).
On the other hand, BAL quasars may differ from ordinary quasars in
some intrinsic fashion, perhaps having a different mean ratio of luminosity
to Eddington luminosity.  Alternatively, we may view them from
a special angle.  If the optical continuum is generated in
an accretion disk, one might expect both that the continuum shape would
depend on $L/L_E$ and that there is wavelength-dependent limb-darkening
(e.g., \citet{H00}).  The latter effect would create a systematic
color offset if the absorbing matter lies in a special direction relative
to the disk.

    That BAL quasars are preferentially redder than ordinary quasars
affects our ability to find them.  The reason our BAL fraction is
larger than the fraction in the LBQS may be that quasars in the LBQS
were selected in part on the basis of blue colors.  If so, the comparative
lack of color bias in the SDSS may be critical for obtaining a fair
estimate of the size and character of the BAL population.  In addition,
if the redder colors are also associated with continuum flux that is
weaker in the BAL direction, the fraction of the sky
around the nucleus covered by BAL material would be larger than
the population fraction of BALs \citep{G97,KV98}.

\subsection{Radio-loud fraction}

    We close with a brief comment about the radio properties of these
BAL quasars.  \citet{W91} found that none of their BALs was radio-loud
and therefore suggested an anti-correlation between the
two properties.  On the other hand, \citet{B00} argued that, if
anything, BAL quasars were {\it more} likely to be found in radio-selected
than in optically-selected samples.

     Radio-loudness is often defined as
$R \equiv F_\nu (5$~GHz$)/F_\nu (4400$~\AA$) \gtrsim 10$ -- 30.   This
criterion can be applied to only about 4\% of the SDSS quasars because
radio data are available for only those brighter at 1.4~GHz than
the FIRST flux limit.  Given the optical flux limit of the SDSS
quasar sample, essentially all those quasars in the EDR detected
by FIRST are radio-loud by this criterion\footnote{Comparing this
radio-loud fraction to the $\simeq 15\%$ found by \citet{K89} in
the PG sample suggests that many radio-loud quasars in the SDSS
are a little too faint to have been detected by FIRST.  \citet{I02}
estimate a radio-loud fraction of $\simeq 8\%$; relative to this
fraction, there are still numerous radio-loud quasars in this
sample that must fall just below the FIRST detection limit.}.  Five of
the 116 BALs found
in our sample are radio-loud by this definition.  Our results are
therefore consistent with the proposition that there is no difference
between the radio-loud BAL fraction and the radio-loud fraction among
ordinary quasars.  However, in view of the very small number of objects
and the incompleteness of radio data for our sample, this conclusion
must be tentative at best.  
\acknowledgments

   We thank Tim Heckman, Gordon Richards, and Tim Reichard for numerous
helpful conversations and suggestions.  J.H.K. was partially supported
by NASA grant NAG5-9187.

\end{document}